\documentclass{article}

\usepackage[numbers]{natbib}

     \usepackage[preprint]{neurips_2020}

\usepackage[utf8]{inputenc} 
\usepackage[T1]{fontenc}    
\usepackage{hyperref}       
\usepackage{url}            
\usepackage{booktabs}       
\usepackage{amsfonts}       
\usepackage{nicefrac}       
\usepackage{microtype}      

\usepackage{amsmath}
\usepackage{amssymb}
\usepackage{subcaption}
\usepackage{wrapfig}
\usepackage{algorithmic}
\usepackage{algorithm}

\usepackage{amsmath,mathtools}

\newcommand{\ro}[1]{%
  \xrightarrow{\mathmakebox[\rowidth]{#1}}%
}
\newlength{\rowidth}
\AtBeginDocument{\setlength{\rowidth}{3.7em}}

\title{fMRI Multiple Missing Values Imputation Regularized by a Recurrent Denoiser}

\author{%
  David Calhas \\
  INESC-ID\\
  Lisbon, Portugal \\
  \texttt{david.calhas@tecnico.ulisboa.pt} \\
  \And
  Rui Henriques \\
  INESC-ID \\
  Lisbon, Portugal \\
  \texttt{rmch@tecnico.ulisboa.pt} \\
}

\begin{document}

\maketitle

\begin{abstract}
Functional Magnetic Resonance Imaging (fMRI) is a neuroimaging technique with pivotal importance due to its scientific and clinical applications. As with any widely used imaging modality, there is a need to ensure the quality of the same, with missing values being highly frequent due to the presence of artifacts or sub-optimal imaging resolutions. Our work focus on missing values imputation on multivariate signal data. To do so, a new imputation method is proposed consisting on two major steps: spatial-dependent signal imputation and time-dependent regularization of the imputed signal. A novel layer, to be used in deep learning architectures, is proposed in this work, bringing back the concept of chained equations for multiple imputation \cite{white2011multiple}. Finally, a recurrent layer is applied to tune the signal, such that it captures its true patterns. Both operations yield an improved robustness against state-of-the-art alternatives. The code is made available on Github.
\end{abstract}

\section{Introduction}\label{section:introduction}

The ability to learn from functional Magnetic Resonance Imaging (fMRI) data is generally hampered by the presence of artifact and limits on the available instrumentation and acquisition protocol, resulting in pronounced missingness. As MRI is collected in frequency space with the usual type of missing/corrupted values occurring at the frequency space. On the other hand, low-quality (voxel space) recordings prevents whole-brain analyzes in clinical settings and is specially pervasive among stimuli-inducing setups in research settings. 
Imputation of incomplete/noisy recordings is critical to classification \cite{smieja2018nnmissing}, synthesis and enhancement tasks. For instance, given the unique spatial and temporal nature of each neuroimaging modality, 
synthesis between distinct modalities is a difficult task (particularly of multivariate time series nature), being imputation an important step of the process \cite{Tran_2017_CVPR,multimodaldiscovery}. Finally, the integration of heterogeneous sources of fMRI recordings by multiple initiatives worldwide also drives the need to increase image resolutions and correct differences arriving from distinct setups. 

This work reclaims the importance of a machine learning based model to perform imputation of multivariate signal data, as opposed to individual-specific imputation. In this context, we propose a novel layer that perform feature based imputation with the principle of chained equations \cite{white2011multiple}. Further, a recurrent layer is proposed to serve as a denoiser to the spatially imputed signal. This two-step principled approach for imputation is illustrated in Fig.\ref{figure:comparison_spatial_reg_approach_theory}. 
Results on resting-state and stimuli-based fMRI recordings validate the robustness of the proposed approach against competitive alternatives. 

This work is organized as follows; Section \ref{section:problem_setting} introduces essential background; Section \ref{section:approach} describes the proposed approach; Section \ref{section:related_work} surveys state-of-the-art work on multivariate time series imputation; Section \ref{section:experiments} presents the experimental setup; Section \ref{section:discussion} discusses the results and places final remarks.

\section{Problem Setting}\label{section:problem_setting}

Multivariate Time Series (MTS) Missing Value Imputation is the focus of this work, specifically high-dimensional MTS data from fMRI recordings. 
The problem is divided into two parts: spatial imputation, in which missing values are sequentially predicted from the existing features; and time dimension regularization, where the imputed values are time tuned.


Consider an fMRI recording to be a multivariate time series $\mathbf{x} \in \mathbb{R}^{v \times t}$, being $v = (v_0, ..., v_{V-1})$ the voxel dimension with $V$ voxels and $t = (t_0, ..., t_{T-1})$ the temporal dimension with $T$ timesteps. An fMRI volume, at timestep $t$, is denoted as $\mathbf{x}^t = [\mathbf{x}^{t}_{0}, ..., \mathbf{x}^{t}_{V-1}] $ and a voxel time series, at voxel $v$, is denoted as $\mathbf{x}_v = [\mathbf{x}^{0}_{v}, ..., \mathbf{x}^{T-1}_{v}] $, each can be seen as the column and row of matrix $\mathbf{x}$, respectively.

In the problem of imputation, consider $\psi \in  \mathbb{R}^{v \times t} \cup \{nan\}$, $\varphi \in  \mathbb{R}^{v \times t}$ and $\mu \in \{0,1\}^{v \times t}$ as the variables involved in the learning phase. The $nan$ symbol denotes a missing value in the fMRI instance $\psi$. $\mu$ is the mask, with $0$ representing a complete value and $1$ a missing value. $\varphi$ is the complete fMRI instance. Illustrating, 
given $V=5$ and $T=7$,
\vskip -0.3cm
\begin{align}
\footnotesize
\label{equation:instance_input}
    \psi= \begin{bmatrix}
        1 & 2 & 3 & 4 & 5 & 6 & 7 \\
        nan & nan & nan & nan & nan & nan & nan\\
        15 & 16 & 17 & 18 & 19 & 20 & 21 \\
        nan & nan & nan & nan & nan & nan & nan \\
        nan & nan & nan & nan & nan & nan & nan
    \end{bmatrix},\hspace{0.25cm} &
\small
    \mu= \begin{bmatrix}
        0 & 0 & 0 & 0 & 0 & 0 & 0 \\
        1 & 1 & 1 & 1 & 1 & 1 & 1 \\
        0 & 0 & 0 & 0 & 0 & 0 & 0 \\
        1 & 1 & 1 & 1 & 1 & 1 & 1 \\
        1 & 1 & 1 & 1 & 1 & 1 & 1
    \end{bmatrix},
\end{align}

\vskip -0.1cm

\begin{equation}
\label{equation:instance_target}
\footnotesize\text{and\hspace{0.2cm}}
    \varphi= \begin{bmatrix}
        1 & 2 & 3 & 4 & 5 & 6 & 7\\
        8 & 9 & 10 & 11 & 12 & 13 & 14\\
        15 & 16 & 17 & 18 & 19 & 20 & 21\\
        22 & 23 & 24 & 25 & 26 & 27 & 28 \\
        29 & 30 & 31 & 32 & 33 & 34 & 35
    \end{bmatrix}.
\end{equation}

Consider the imputation of missing values is made by a model, $I$, and each imputed value is denoted as $\iota_{v}^{t} \in \mathbb{R}$. Continuing our example, 
an fMRI instance, $\psi$, after processing by $I$, is $\iota \in \mathbb{R}^{v \times t}$,

\vskip -0.3cm
\begin{equation}\label{equation:instance_imputed}
\footnotesize
    I(\psi) = \iota= \begin{bmatrix}
        1 & 2 & 3 & 4 & 5 & 6 & 7\\
        \iota_{1}^{0} & \iota_{1}^{1} & \iota_{1}^{2} & \iota_{1}^{3} & \iota_{1}^{4} & \iota_{1}^{5} & \iota_{1}^{6}\\
        15 & 16 & 17 & 18 & 19 & 20 & 21 \\
        \iota_{3}^{0} & \iota_{3}^{1} & \iota_{3}^{2} & \iota_{3}^{3} & \iota_{3}^{4} & \iota_{3}^{5} & \iota_{3}^{6} \\
        \iota_{4}^{0} & \iota_{4}^{1} & \iota_{4}^{2} & \iota_{4}^{3} & \iota_{4}^{4} & \iota_{4}^{5} & \iota_{4}^{6}
    \end{bmatrix}.
\end{equation}

\footnote{Please note that values in (\ref{equation:instance_input}) to (\ref{equation:instance_imputed}) are shown for simplicity sake, not resembling real fMRI values.}

Considering typical fMRI resolution, each voxel has a 3D euclidean point correspondence. As such, a spatial distance matrix, $d \in \mathbb{R}
^{V \times V}$, is defined, where $d_{i,j}$ corresponds to the distance between voxels $v_i$ and $v_j$, with $i,j \in \{0, ..., V-1\} \in \mathbb{N}$.

Missing voxels can occur at random or, in contrast, be spatially autocorrelated within a variable number of regions, resembling the characteristics of an artifact. Both modes are targeted in this work.

\section{Proposed Approach}\label{section:approach}

In this section, two main steps are proposed to perform imputation of missing values from MTS data: 

\begin{itemize}
    \item \textbf{Spatial Imputation}, where imputation is done by estimating missing values, $\mu_{i}^{j} = 1$, from complete features, $\mu_{k}^{l} = 0$, with $i,k \in \{0,...,V-1\}, j,l \in \{0,...,T-1\}$.
    \item \textbf{Time Series Regularization}, where the previously derived missing values are processed by a recurrent neural network.
\end{itemize}

This two-step approach is shown in section \ref{section:discussion} to outperform competitive baselines. 

\subsection{Spatial Imputation}\label{subsection:approach_spatial}

We propose a novel neural network layer, $\Phi$, that performs imputation inspired by the chained equations principle proposed by \citet{white2011multiple}. This imputation method consists of filling a missing value at a time, and using its estimate to guide the imputation of the remaining missing values. 

The priority in which the values are filled is given by the pairwise voxel correlation matrix, $C$, computed from the training set data.

This layer, $\Phi$, is characterized by a weight matrix $W_\Phi$ and bias $B_\Phi$,
\begin{align}\label{equation:layer_weights}~
\small
    W_\Phi = \begin{bmatrix}
        0 & w^{1}_{0} & ... & w^{V-1}_{0} \\
        w^{0}_{1} & 0 & ... & w^{V-1}_{1} \\
        \vdots & \vdots & \ddots & \vdots \\
        w^{0}_{V-1} & w^{1}_{V-1} & ... & 0 \\
    \end{bmatrix} ,\hspace{0.2cm} &
    B_\Phi = \begin{bmatrix}
        b_{0} \\
        b_{1} \\
        \vdots\\
        b_{V-1}
    \end{bmatrix}.
\end{align}

The activation function of $\Phi$ is linear, making the imputed values a linear combination of the already filled and complete values. 
The weight matrix has a zero-filled diagonal for each voxel to be described as a linear combination of the remaining voxels.

Since this neural function estimates a single value at a time, one only needs to compute the dot product of the missing value, $v$, with the corresponding column, $W^{v}_{\Phi}$, and add the bias, $B^{v}_{\Phi}$. The imputation operation of a missing value, $v$, is denoted as\hspace{0.1cm} $\ro{\psi^{0} {W_\Phi}^{v} + {B_\Phi}^{v}}$.

Let us consider the input presented in Equation \ref{equation:instance_input} (section \ref{section:problem_setting}), with $\psi^{0} = [1, nan, 15, nan, nan]$, and\hspace{0.1cm} $\ro{\psi^{0} {W_\Phi}^{c} + {B_\Phi}^{c}}$ as the operation made by layer $\Phi$ at each iteration to impute a missing value, $c$. Being $c = max(C_{\mu^{0} = 1})$ the missing voxel that has the highest correlation with the complete and filled voxels. 
This scheme allows an imputation of missings under the chained equation principle,

\begin{equation}\label{equation:chained_imputation}
    \Phi(\psi^{0}) = \begin{bmatrix}
           1 \\
           nan \\
           15 \\
           nan \\
           nan \\
      \end{bmatrix}^T \ro{\psi^{0} {W_\Phi}^{c} + {B_\Phi}^{c}} \begin{bmatrix}
           1 \\
           \phi_1 \\
           15 \\
           nan \\
           nan \\
      \end{bmatrix}^T \ro{\psi^{0} {W_\Phi}^{c} + {B_\Phi}^{c}} \begin{bmatrix}
           1 \\
           \phi_1 \\
           15 \\
           nan \\
           \phi_4 \\
      \end{bmatrix}^T \ro{\psi^{0} {W_\Phi}^{c} + {B_\Phi}^{c}} \begin{bmatrix}
           1 \\
           \phi_1 \\
           15 \\
           \phi_3 \\
           \phi_4 \\
      \end{bmatrix}^T = \phi .
\end{equation}

\begin{wrapfigure}{L}{0.4\textwidth}
    \begin{minipage}[t]{\linewidth}
        \vspace{-0.8cm}
        \begin{algorithm}[H]
            \caption{\footnotesize{$\Phi$ chained imputation cycle}\hspace{-0.3cm}}
            \label{algorithm:chained_imputation}
            \begin{algorithmic}
                \STATE $\phi \leftarrow \psi^t$
                \WHILE{$\sum {\mu^t} > 0$}
                    \STATE $c \leftarrow max(C_{\mu^t=1})$
                    \STATE $\phi_{c} \leftarrow \psi^{t} {W_\Phi}^{c} + {B_\Phi}^{c}$
                    \STATE $\mu^{t}_{c} \leftarrow 0$
                \ENDWHILE
                \STATE $\phi$
            \end{algorithmic}
        \end{algorithm}
        \vspace{-1.2cm}
    \end{minipage}
\end{wrapfigure}

$\phi$ is the output of layer $\Phi$, with $\phi_i$ corresponding to the $v_i$ missing voxel imputation. There is a total of $\sum {\mu^t}$ iterations equal to the number of missing values. Algorithm \ref{algorithm:chained_imputation} presents the pseudocode for this imputation scheme.

This layer imputes all missing values for every time frame, $t$, of an fMRI recording, $\psi^t$. Imputation is merely done accounting other features, therefore spatial.

This layer contrasts with the traditional dropout layer for imputing missing values. In a dropout layer, each weights' column, $W^{v}$, shows intra-correlation, converging to a single target independently. However, there is no inter-column correlation/dependency. $\Phi$ layer forces the columns, $W_{\Phi}^{v}$, to be inter-correlated, converging to the same target as a unit. Here, the estimates of a column (the imputed values) influence the estimates of the upcoming columns along the imputation process.

\subsection{Time Dimension Regularization}\label{subsection:approach_time}

\begin{figure}[ht]
    \centering
    \caption{\small{Time regularization (denoising) component.}}
    \vspace{0cm}
    \begin{subfigure}{.45\textwidth}
        \caption{\small Error between the true signal and\\ \mbox{spatially imputed signal}}
        \includegraphics[width=\textwidth,height=4cm]{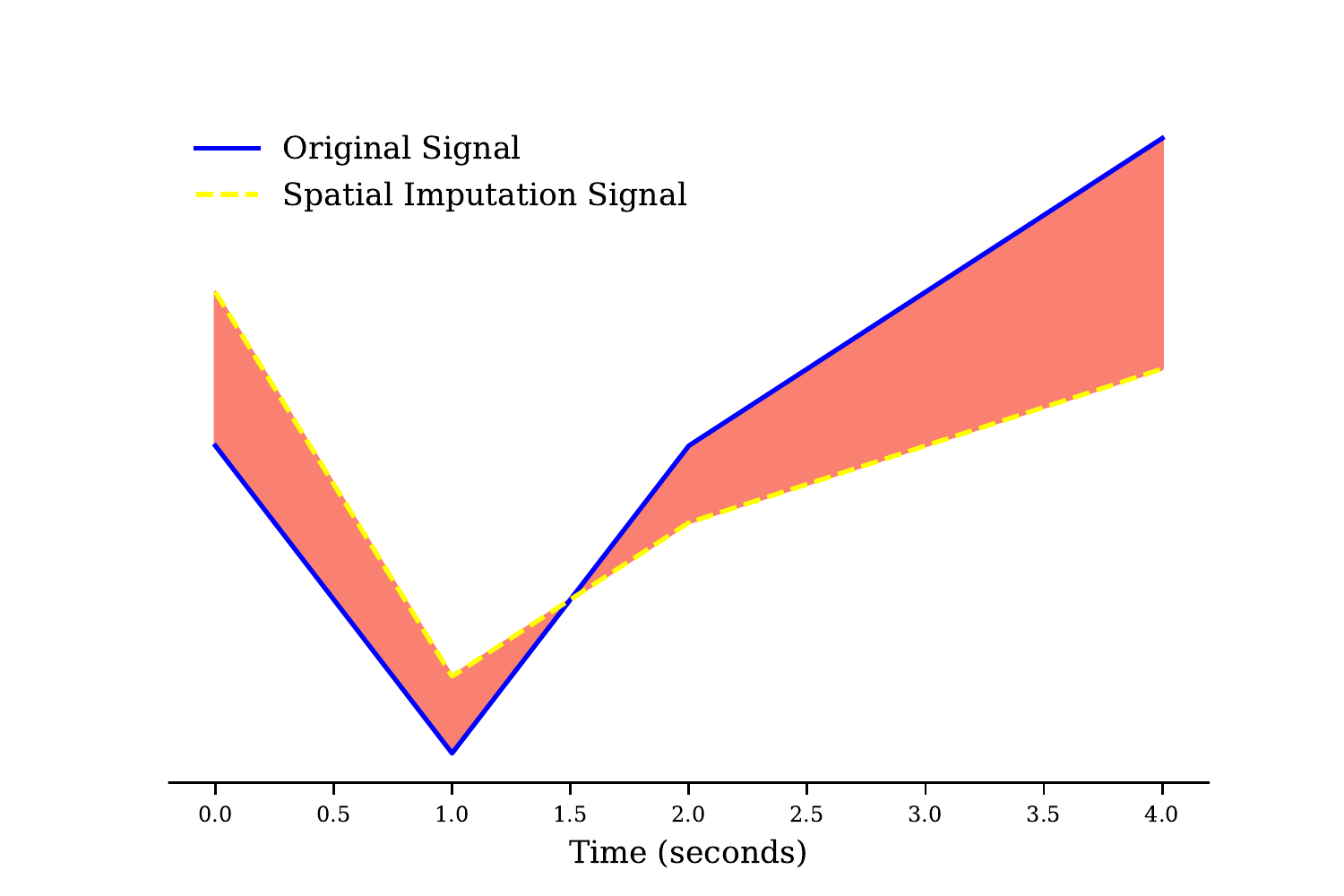}
        \label{fig:spatial_error_theory}
    \end{subfigure}
    \begin{subfigure}{.45\textwidth}
        \caption{\small Error between the true signal and spatially\\ \mbox{imputed signal after time regularization}}
        \includegraphics[width=\textwidth,height=4cm]{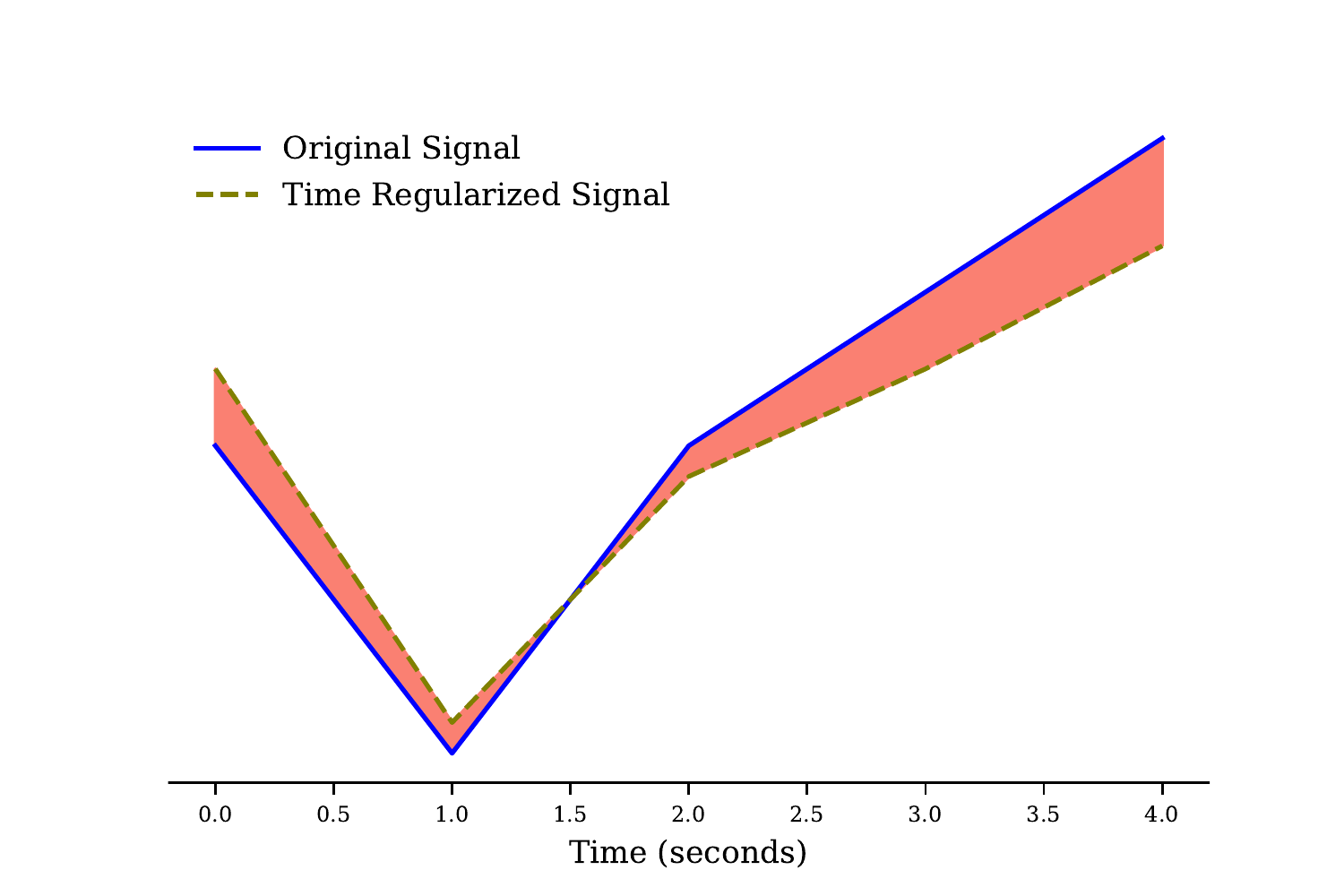}
        \label{fig:time_reg_error_theory}
    \end{subfigure}
    \vspace{-1cm}
    \label{figure:comparison_spatial_reg_approach_theory}
\end{figure}

Once spatial imputation is done for each voxel, $v$, of an fMRI recording, $\phi_v$, the imputed values are fed to a recurrent layer, tweaking the signal in such a way that it emulates the target time series patterns. The recurrent layer removes the noise created by the spatial imputation method. We refer to this recurrent layer as the Denoiser, $D$, component of the imputation pipeline. An illustration of this noise removal is shown in Figure \ref{figure:comparison_spatial_reg_approach_theory}.

$D$ is a single layer Gated Recurrent Unit (GRU) \cite{cho2014learning}. This choice was motivated by results collected against its rival Long-Short Term Memory Layer on the target task, and further supported by studies showing that GRU performs well on datasets with limited observations \cite{lu2017bidirectional}, the common case when learning Neuroimaging datasets. \citet{imp2018luo, che2018recurrent} altered the internal function of a GRU layer to perform direct imputation on a multivariate time series. In contrast, we maintain the GRU layer as it was originally proposed \cite{cho2014learning} since our purpose is to remove noise and capture the desired temporal patterning properties of the (neurophysiological) signal.

The imputation model we propose is denoted as $\Phi+D$, corresponding to the junction of the spatial imputation scheme described in Section \ref{subsection:approach_spatial} and the time regularization described in section \ref{subsection:approach_time}.

\subsection{Validation and Hyperparameters}\label{subsection:validation}

Bayesian Optimization (BO) \cite{snoek2012practical} was used to find the hyperparameters that best fitted the pipeline. For each BO iteration a $6$-Fold Cross Validation was ran and the Mean Absolute Error (MAE) was the metric targeted as alternative residue-based scores can overly focus on the minimization or large residues. Manual tweaking was performed before to check which optimizers should be used. In consensus, Adam optimizer \cite{kingma2014adam} is used to optimize the $D$ and $\Phi$ trainable parameters. Missing data generation (explained in section \ref{section:random_removal}) is made at every iteration of the BO algorithm, in order to avoid overfitting towards a certain missing values setting.

Besides $\Phi+D$, BO was also ran for $\Phi$, Dropout and Dropout$+D$. 
The hyperparameters 
were subjected to 
a total of $50$ iterations. Their range spaces are: $\Phi$ learning rate, $L_\Phi \in [1\mathrm{e}{-5},1\mathrm{e}{-2}] \in \mathbb{R}$; $D$ learning rate, $L_D \in [1\mathrm{e}{-5},1\mathrm{e}{-2}]  \in \mathbb{R}$; $\Phi$ number of epochs, $E_\Phi \in \{2,3,4,5\} \in \mathbb{N}$; $D$ number of epochs, $E_D \in \{2,3,4,5\}  \in \mathbb{N}$; number of alternating epochs, $E \in \{2,4,8,10\}  \in \mathbb{N}$; $D$ L1-norm regularization constant, $R_D \in [1\mathrm{e}{-5}, 3]  \in \mathbb{R}$; $D$ Use of bias, $B_D \in \{0, 1\}  \in \mathbb{N}$; $D$ Dropout \cite{srivastava2014dropout}, $Dr_D \in [0, 3\mathrm{e}{-1}]  \in \mathbb{R}$; $D$ Recurrent Dropout \cite{srivastava2014dropout}, $RDr_D \in [0, 3\mathrm{e}{-1}]  \in \mathbb{R}$.

\section{Related Work}\label{section:related_work}

In this section, state-of-the-art contribution on multivariate time series imputation are presented, discussed and 
comparing against our work.

\citet{smieja2018nnmissing} handles missing values by training the parameters of a Gaussian Mixture Model (GMM) along with a neural network. Missing values are imputed at the first hidden layer by computing the expected activation of neurons (instead of just calculating the expected input). Thus the imputation is not made by single values, instead it is modeled by a GMM. Although competitive, this approach 
performed worse than a Context Encoder (CE) \cite{Pathak_2016_CVPR} which, in contrast, learns from a loss function using the complete data. In contrast to this work, we perform chained imputation based on correlation ranking, instead of imputing values at one step by taking advantage of a GMM for each feature.
\citet{che2018recurrent} proposed a variant of the Gated Recurrent Unit (GRU) to handle generic time series with missing values, claiming that, by placing masks and time intervals in accordance with the properties of missing patterns, their model, GRU-D, is able to take advantage of missing data to improve classification. Masking and time intervals in GRU-D \citep{che2018recurrent} are represented using a decay term computed by a exponentiated negative rectifier function. Given a missing occurrence, the decay at that timestep is used over time to converge to the empirical mean.
\citet{cao2018brits} performs missing value imputation from two estimates produced from a spatial-based and a recurrent model. Results on air quality, health care and human activity datasets show superiority among baselines reaching $11.56$, $0.278$ and $0.219$ MAE. The imputation task is mapped as a classification task to learn the target models. 
After imputation is made separately by these two models, a linear combination, defined by a parameter, is computed to produce the final estimate. In contrast, we take advantage of a recurrent model described in section \ref{subsection:approach_time} to remove prediction noise from the spatial imputation model and strengthen the temporal consistency. Further, \cite{cao2018brits} assume missing values occur sporadically in a feature time series of the multivariate time series, on the other hand our work focuses on the imputation of whole feature time series to resemble the characteristics missing neuroimaging data. 
\citet{luo2018multivariate} propose a disruptive model based on a Generative Adversarial Architecture \cite{goodfellow2014generative}. The Discriminator and Generator components are both an internally tweaked version of GRU. The Discriminator classifies generated and real multivariate time series samples and the Generator performs imputation on samples with missing values. Results show classification superiority using the AUC metric.
\citet{fortuin2019gp} uses a Variational Autoencoder Architecture \cite{kingma2013auto} to perform imputation. The Generative model is a Gaussian Process that generates samples from complete encoded feature representations.

All the discussed works in this section perform multivariate time series imputation from incomplete data. For a more objective assessment, our work tests imputation methods over complete datasets with generated missing entries and regions according to the proposed validation scheme. 

\section{Experimental Setting}\label{section:experiments}

This section describes the setting in which results were gathered. The baselines, target datasets, missing generation procedures, and metrics are detailed in sections \ref{subsection:baselines}, \ref{subsection:datasets}, \ref{section:random_removal} and \ref{subsection:evaluation_metrics}, respectively.

\subsection{Baselines}\label{subsection:baselines}

For the sake of comparison, the following baselines were implemented to gather the results:

\begin{itemize}
    \item kNN imputation \cite{hastie01statisticallearning}
    \item Barycenter \cite{petitjean2011barycenter}
    \item MICE \cite{white2011multiple}
    \item Mean imputation
    \item Context Encoder (CE) \cite{Pathak_2016_CVPR}
    \item Dropout
    \item Dropout with time regularization, denoted as Dropout$+D$
    \item $\Phi$ with no time regularization
\end{itemize}

kNN was used with a $k$=3 since there were no overall significant improvements for alternative $k$. 
Barycenter computes the average time series of a multivariate time series and imputation is made with the average time series under the Dynamic Time Wrapping (DTW) distance criterion. MICE, a.k.a. multiple imputation by chained equations, is a method similar to ours as it has its basis on the same rationale of imputing one missing value at a time. It is thus considered a baseline as well. Mean imputation method takes the mean of each feature from the training set and performs kNN imputation ($k=3$) if there is no information about a voxel in the training set. Context Encoder (CE) is a simple Autoencoder with 2-Dimensional Convolution Layers. Dropout method drops the weights linked to the missing values. Dropout was also extended with the time regularization scheme presented in section \ref{subsection:approach_time}. Finally, we also considered comparing $\Phi$ alone against $\Phi+D$ to measure the impact of reshaping the imputed time series. 

\subsection{Datasets}\label{subsection:datasets}

\noindent\textbf{EEG, fMRI and NODDI Dataset}.\label{subsubsection:noddi_dataset} 
This dataset \cite{deligiani2014dataset, deligianni2016noddidataset} contains $16$ individuals, with an average age $32.84 \pm 8.13$ years. Simultaneous EEG-fMRI recordings of resting state with eyes open (fixating a point) were acquired. 
The fMRI acquisition was done using a T2-weighted gradient-echo EPI sequence with: $300$ volumes, TR of $2160$ ms, TE of $30$ ms, $30$ slices with $3.0$ millimeters (mm), 
voxel size of $3.3\times3.3\times4.0$ mm and a field of view of $210\times210\times120$ mm. For a more detailed description please see the dataset references \cite{deligiani2014dataset, deligianni2016noddidataset}. The dataset is available for download in its original source at \url{https://osf.io/94c5t/}. Each individual recording was divided into $24$ equally sized time series of fMRI volumes. Each time series is $28$ seconds long and resampled to a $2$ second period. The \textit{training set} is composed of $12$ individuals and the \textit{test set} of $4$ individuals.
\vskip 0.2cm 

\noindent\textbf{Auditory and Visual Oddball EEG-fMRI Dataset}.\label{subsubsection:auditory_dataset} 
This dataset \cite{walz2013simultaneous02dataset,walz2014simultaneous02dataset,conroy2013fast02dataset} contains $17$ individuals. Simultaneous EEG-fMRI recordings were performed while the subjects laid down. Stimuli of auditory and visual nature were given to the subjects, which makes this a stimuli-based dataset. The fMRI imaging acquisition was made with a 3T Philips Achieva MR Scanner with: single channel send and receive head coil, EPI sequence, $170$ TRs per run with a TR of $2000$ ms and $25$ ms TE, $170$ TRs per run with a $3\times3\times4$ mm voxel size and $32$ slices with no slice gap. For a more detailed description of the dataset please refer to \cite{walz2013simultaneous02dataset}. The dataset is available for download in its original source at \url{https://legacy.openfmri.org/dataset/ds000116/}. Each individual recording was divided into $12$ equally sized time series of fMRI volumes. Each time series is $28$ seconds long, sampled at $2$ seconds period. The \textit{training set} is composed of $12$ individuals and the \textit{test set} $5$ individuals.
\vskip 0.2cm

In both datasets, the $6$-Fold Cross Validation schema explained in section \ref{subsection:validation}, is structured by a folding with sets of $10$ and $2$ individuals, for training and \textit{validation}, respectively. One might argue that because the dataset contains multiple individuals and not a single subject on the same scanner, it might be difficult to fit the correlation matrix, due to different line ups and brain sizes. To tackle it, we downsample the fMRI spatial resolution by a factor of $6$, going from $\approx30$K voxels in total to $100$ voxels to represent the whole brain of multiple subjects. The results in section \ref{section:experiments}, show the feasibility of the task and support the claim.
The selected datasets offer an important basis to validate the target imputation models due to their artifact susceptibility caused by the instrumentation (simultaneous EEG-fMRI recording) and monitoring protocol.

\subsection{Random Value and Region Removal}\label{section:random_removal}

\begin{wrapfigure}{R}{0.55\textwidth}
    \begin{minipage}[t]{\linewidth}
        \vspace{-0.8cm}
\begin{algorithm}[H]
\small
    \caption{\small Random Region Removal}
    \label{algorithm:region_removal}
    \begin{algorithmic}
        \STATE $R \leftarrow r \times V$
        \STATE $\text{removed} \leftarrow 0$
        \STATE $v_{to\_remove} = \text{random\_choice}(v)$
        \WHILE{$\text{removed} < R$}
            \STATE $\text{remove}(v_{to\_remove})$
            \STATE $\text{removed} \leftarrow \text{removed} + 1$
            \STATE $\text{adjacent\_voxels} \leftarrow d_v \leq 1.0$
            \WHILE{$\text{random\_choice}(\text{adjacent\_voxels}) \wedge \text{removed} < R$}
                \STATE $v_{to\_remove} \leftarrow \text{random\_choice}(\text{adjacent\_voxels})$
                \STATE $\text{remove}(v_{to\_remove})$
                \STATE $\text{removed} \leftarrow \text{removed} + 1$
                \STATE $\text{adjacent\_voxels} \leftarrow \text{adjacent\_voxels} \cup d_v \leq 1.0$
            \ENDWHILE
        \ENDWHILE
    \end{algorithmic}
\end{algorithm}
        \vspace{-0.7cm}
    \end{minipage}
\end{wrapfigure}

This manuscript performs missing data imputation using a supervised learning method. To guarantee an objective assessment, 
we operate on a complete dataset \cite{deligiani2014dataset, deligianni2016noddidataset} and generate missing data using two distinct procedures: random value removal and random region removal, where the last captures the spatially correlated nature of artifacts. 
Random region removal can be further used to assess the applicability of supervised principles of imputation to facilitate the synthesis of images (e.g. EEG-to-fMRI). 
The occurrence of a missing on a voxel from a certain fMRI instance generally implies the absence of all values for that voxel along the time axis, $t=(0, ..., T-1)$. 
As such, We do not consider differentiated random removal across time frames (i.e. removing a different set of voxels for each time frame). 
The random value removal strategy generates missings from uniform space distribution, while random region removal strategy is outlined in Algorithm \ref{algorithm:region_removal}. 
To remove random regions, a value is chosen at random and removed along with its adjacent values. The number of adjacent values is set by a removal rate, $r$, indicating the number of values to remove per region or, in alternative, by a maximum value of adjacent values. 
Adjacencies are identified from the 3-Dimensional fMRI voxel coordinates. 



\subsection{Evaluation Metrics}\label{subsection:evaluation_metrics}

For comparing results, two metrics are computed -- Mean Absolute Error (MAE) and Root Mean Squared Error (RMSE) -- and differences on these residues statistically tested. Consider $N$ to be the individuals, $S$ the number of recordings per individual, 
and $M$ the total number of missing values/voxels. In this context,
\vskip -0.3cm
\begin{equation}\label{equation:mae}
\small
    \text{MAE }_i = { \frac{\sum_{v=0}^{M} {d(\iota_{v},\hat{\iota}_{v})}} {M} } \hspace{0.2cm}\text{ and }    \text{ RMSE }_i = \sqrt{ \frac{\sum_{v=0}^{M} {d(y_{v},\hat{y}_{v})^2}} {M} },
\end{equation}

where $\hat{\iota}_{v} = I(\psi_v)$ is the estimate of the missing time series $\iota_{v}$ associated with voxel $v$, and $d$ is the Manhattan distance (the sum of absolute-valued residues along time). The final errors, $e$, are then averaged across all available recordings, 
\vskip -0.3cm
\begin{equation}
\small
e = \frac{1}{N\times S}\sum_{i=0}^{N\times S} \text{error }_i\hspace{0.1cm}.
\end{equation}


\section{Experiment Results}\label{section:discussion}

Tables \ref{table:mae_results_time} to \ref{table:second_rmse_results_time} present the gathered results from assessing the imputation methods (section \ref{subsection:baselines}) over the target datasets (section \ref{subsection:datasets}) using the random region removal strategy (section \ref{section:random_removal}). The results outline the relevance of applying a recurrent layer D for the time-sensitive regularization of spatially imputed signals in comparison with Context Encoder and MICE alternatives. MICE does not scale with increases on missing rate, due to its inability to deal with features that have not been observed before. The chained imputation principle ($\Phi+D$) further shows slight improvements against weakly-correlated signals under a Dropout$+D$ architecture, indicating the importance of identifying voxel priorities. 
Considering the general performance limits of $k$NN ($k$=3) and DTW-based barycenter, the results further motivate the difficulty of the task at hands and underline the role of learning from the available data instances in light of the . 

Results gathered using the alternative random value removal strategy (section \ref{section:random_removal}) are provided in \textit{Appendix A}. Generally, these appended results yield similar ranks among the compared methods. 

Please also refer to \textit{Appendix B} for results collected using alternative residue-based scores that offer complementary information on the spatial adequacy of the assessed methods, further supporting the relevance of the proposed $\Phi+D$ imputation approach. 

\begin{wrapfigure}{r}{0.54\textwidth}
    \vspace{-0.4cm}
    \centering
    \caption{\small MAE for varying brain volumes on the EEG-fMRI-NODDI testing data under a $50\%$ missing rate.}
    \includegraphics[width=\linewidth]{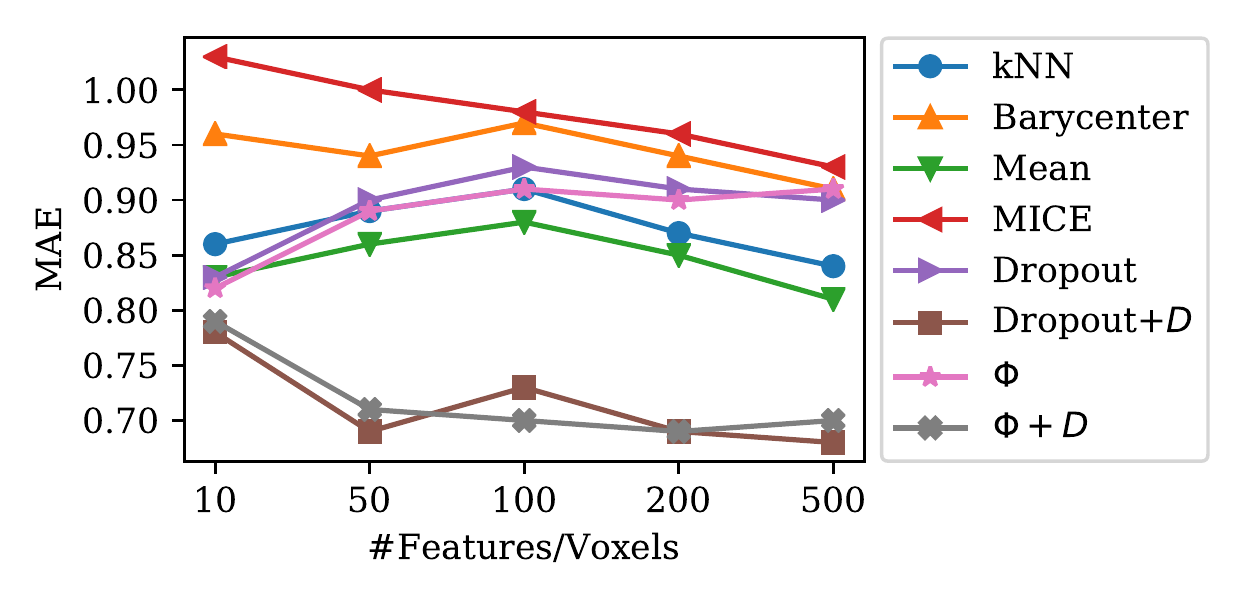}
    \label{fig:performance_scalability}
    \vspace{-1cm}
\end{wrapfigure}
Figure \ref{fig:performance_scalability} provides a complementary view on the performance of imputation methods for the first dataset when considering a varying brain volume under analysis. The gathered results evidence the superiority of the proposed approach and suggest that performance is independent of the spatial extent. 

\begin{figure}[ht]
\vskip -0.3cm
    \centering
    \caption{\small Impact of time regularization on the imputation of voxel with coordinates [14,29,14].} 
    \vspace{0cm}
    \begin{subfigure}{.45\textwidth}
        \caption{Error between original signal and spatially imputed signal using $\Phi$}
        \includegraphics[width=\textwidth,height=3.5cm]{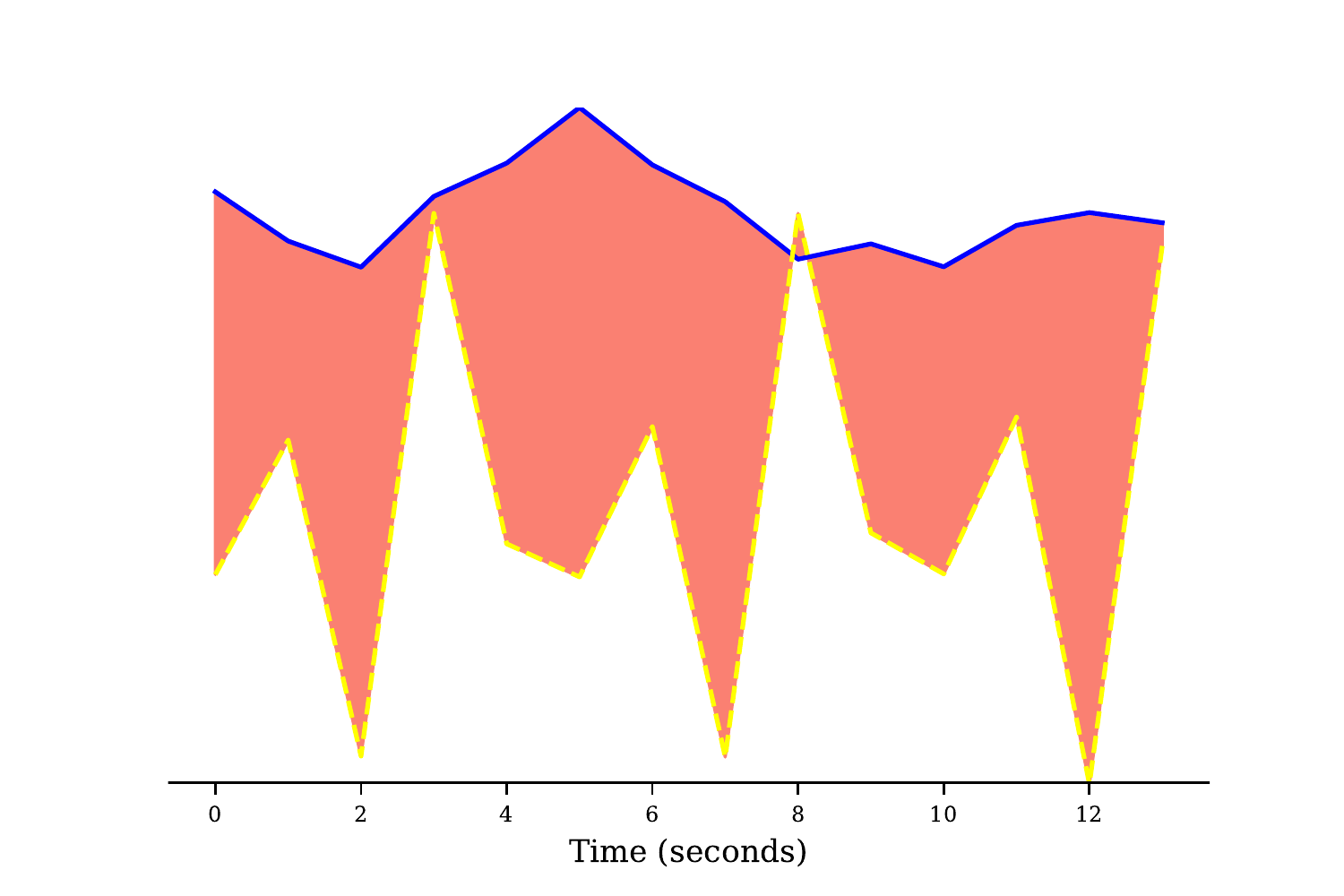}
        \label{fig:spatial_error}
    \end{subfigure}
    \begin{subfigure}{.45\textwidth}
        \caption{Error between original signal and time regularized signal using $\Phi+D$}
        \includegraphics[width=\textwidth,height=3.5cm]{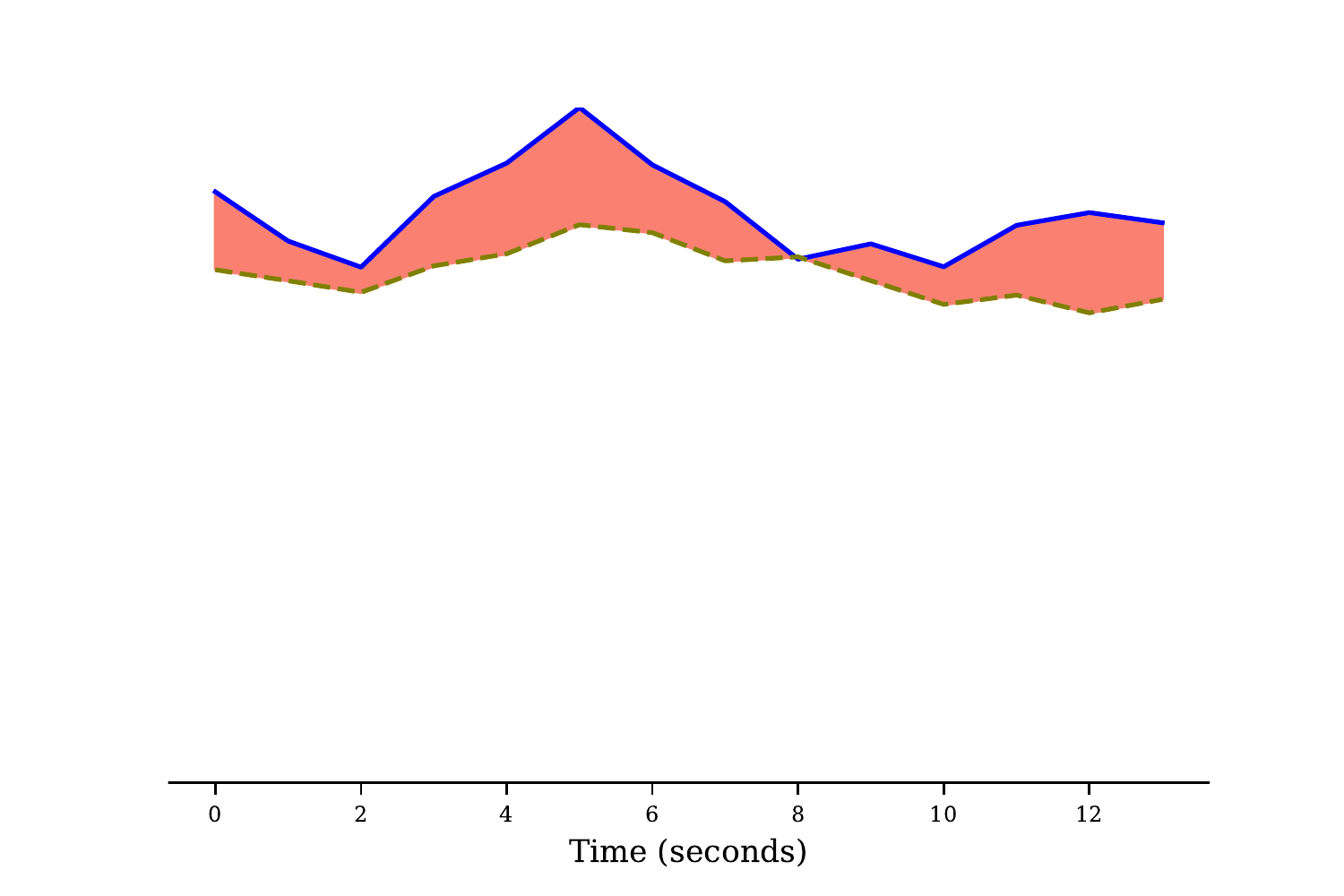}
        \label{fig:time_reg_error}
    \end{subfigure}
    \vspace{-1cm}
    \label{figure:comparison_spatial_reg_approach}
\end{figure}

Figure \ref{figure:comparison_spatial_reg_approach} illustrates the denoising property of $D$ on a randomly selected missing voxel -- with coordinates [14,29,14] -- from the first dataset. It compares, side by side, the error of a single time series imputed spatially, by $\Phi$, with the error of an imputed signal with time regularization, by $\Phi+D$. This image, together with results (Tables \ref{table:mae_results_time}-\ref{table:second_rmse_results_time}), show the importance of this recurrent layer to capture the neurophysiological temporal patterning of the signal. 

\begin{table}[ht]
    \centering
    \caption{MAE on time axis results on the EEG, fMRI and NODDI Test Set.}
    \begin{tabular}{p{2.2cm} p{1.7cm} p{1.7cm} p{1.7cm} p{1.7cm} p{1.7cm}}
        \hline
        Missing & $10\%$ & $25\%$ & $50\%$ & $75\%$ & $90\%$ \\
        \hline
        kNN & $0.69\pm0.73$ & $0.67\pm0.72$ & $0.68\pm0.76$ & $0.66\pm0.77$ & $0.64\pm0.84$ \\
		Barycenter \cite{petitjean2011barycenter} & $0.70\pm0.74$ & $0.69\pm0.74$ & $0.74\pm0.84$ & $0.73\pm0.82$ & $0.78\pm0.86$ \\
		Mean & $0.70\pm0.73$ & $0.68\pm0.72$ & $0.67\pm0.78$ & $\mathbf{0.65}\pm0.77$ & $0.65\pm0.79$ \\
		CE \cite{Pathak_2016_CVPR} & $0.71\pm0.72$ & $0.74\pm0.69$ & $0.76\pm0.74$ & $0.76\pm0.71$ & $0.78\pm0.71$ \\
		MICE \cite{white2011multiple} & $0.71\pm0.74$ & $0.71\pm0.73$ & $0.78\pm0.85$ & $1.36\pm4.42$ & $0.71\pm0.85$ \\
		Dropout & $0.70\pm0.73$ & $0.70\pm0.73$ & $0.71\pm0.77$ & $0.67\pm0.74$ & $\mathbf{0.65}\pm0.76$ \\
		Dropout$+D$ & $\mathbf{0.55}\pm0.59$ & $\mathbf{0.58}\pm0.62$ & $\mathbf{0.63}\pm0.70$ & $\mathbf{0.65}\pm0.69$ & $\mathbf{0.65}\pm0.76$ \\
		$\Phi$ & $0.70\pm0.73$ & $0.70\pm0.73$ & $0.70\pm0.77$ & $\mathbf{0.64}\pm0.77$ & $\mathbf{0.64}\pm0.79$ \\
		$\Phi+D$ & $\mathbf{0.55}\pm0.59$ & $\mathbf{0.57}\pm0.61$ & $\mathbf{0.66}\pm0.76$ & $\mathbf{0.64}\pm0.78$ & $\mathbf{0.65}\pm0.81$ \\
        \hline
    \end{tabular}
    \vspace{-0.5cm}
    \label{table:mae_results_time}
\end{table}

\begin{table}[t!]
    \centering
    \caption{RMSE on time axis results on the EEG, fMRI and NODDI Test Set.}
    \begin{tabular}{p{2.2cm} p{1.7cm} p{1.7cm} p{1.7cm} p{1.7cm} p{1.7cm}}
        \hline
        Missing & $10\%$ & $25\%$ & $50\%$ & $75\%$ & $90\%$ \\
        \hline
        kNN & $1.00\pm0.79$ & $0.99\pm0.79$ & $1.02\pm0.84$ & $1.01\pm0.85$ & $1.05\pm0.93$ \\
		Barycenter \cite{petitjean2011barycenter} & $1.02\pm0.81$ & $1.02\pm0.81$ & $1.12\pm0.92$ & $1.10\pm0.89$ & $1.16\pm0.94$ \\
		Mean & $1.01\pm0.80$ & $0.99\pm0.79$ & $1.03\pm0.86$ & $1.01\pm0.85$ & $1.02\pm0.87$ \\
		CE \cite{Pathak_2016_CVPR} & $1.01\pm0.78$ & $1.01\pm0.74$ & $1.06\pm0.80$ & $1.04\pm0.76$ & $1.06\pm0.76$ \\
		MICE \cite{white2011multiple} & $1.02\pm0.80$ & $1.02\pm0.80$ & $1.15\pm0.93$ & $4.63\pm5.50$ & $1.11\pm0.93$ \\
		Dropout & $1.01\pm0.80$ & $1.02\pm0.80$ & $1.05\pm0.84$ & $1.00\pm0.82$ & $1.00\pm0.84$ \\
		Dropout$+D$ & $\mathbf{0.81}\pm0.65$ & $\mathbf{0.85}\pm0.68$ & $\mathbf{0.94}\pm0.76$ & $\mathbf{0.95}\pm0.75$ & $\mathbf{1.00}\pm0.84$ \\
		$\Phi$ & $1.01\pm0.80$ & $1.01\pm0.80$ & $\mathbf{1.04}\pm0.85$ & $\mathbf{1.00}\pm0.85$ & $\mathbf{1.02}\pm0.88$ \\
		$\Phi+D$ & $\mathbf{0.81}\pm0.65$ & $\mathbf{0.84}\pm0.67$ & $\mathbf{1.00}\pm0.83$ & $\mathbf{1.02}\pm0.87$ & $\mathbf{1.03}\pm0.89$ \\
        \hline
    \end{tabular}
    \label{table:rmse_results_time}
    \vspace{-0.5cm}
\end{table}

\begin{table}[t!]
    \centering
    \caption{MAE on time axis results on the Auditory and Visual Oddball EEG-fMRI Test Set.}
    \begin{tabular}{p{2.2cm} p{1.7cm} p{1.7cm} p{1.7cm} p{1.7cm} p{1.7cm}}
        \hline
        Missing & $10\%$ & $25\%$ & $50\%$ & $75\%$ & $90\%$ \\
        \hline
        kNN & $0.90\pm0.80$ & $0.92\pm0.78$ & $0.91\pm0.77$ & $0.88\pm0.67$ & $0.86\pm0.61$ \\
		Barycenter\cite{petitjean2011barycenter} & $0.91\pm0.82$ & $0.95\pm0.82$ & $0.97\pm0.80$ & $0.97\pm0.76$ & $0.97\pm0.73$ \\
		Mean & $0.92\pm0.78$ & $0.90\pm0.75$ & $0.88\pm0.72$ & $\mathbf{0.84}\pm0.63$ & $\mathbf{0.83}\pm0.58$ \\
		CE \cite{Pathak_2016_CVPR} & $\mathbf{0.89}\pm0.63$ & $\mathbf{0.88}\pm0.63$ & $\mathbf{0.84}\pm0.59$ & $\mathbf{0.84}\pm0.59$ & $\mathbf{0.82}\pm0.56$ \\
		MICE\cite{white2011multiple} & $0.94\pm0.81$ & $0.97\pm0.82$ & $0.98\pm0.81$ & $1.42\pm3.01$ & $3.61\pm16$ \\
		Dropout & $0.93\pm0.81$ & $0.93\pm0.79$ & $0.93\pm0.77$ & $0.88\pm0.67$ & $0.85\pm0.61$ \\
		Dropout$+D$ & $\mathbf{0.58}\pm0.42$ & $\mathbf{0.66}\pm0.47$ & $\mathbf{0.73}\pm0.40$ & $\mathbf{0.75}\pm0.55$ & $0.86\pm0.61$ \\
		$\Phi$ & $\mathbf{0.92}\pm0.81$ & $\mathbf{0.93}\pm0.79$ & $\mathbf{0.91}\pm0.75$ & $\mathbf{0.85}\pm0.64$ & $\mathbf{0.83}\pm0.58$ \\
		$\Phi+D$ & $\mathbf{0.57}\pm0.41$ & $\mathbf{0.65}\pm0.46$ & $\mathbf{0.70}\pm0.52$ & $\mathbf{0.81}\pm0.55$ & $\mathbf{0.82}\pm0.56$ \\
        \hline
    \end{tabular}
    \label{table:second_mae_results_time}
    \vspace{-0.5cm}
\end{table}

\begin{table}[t!]
    \centering
    \caption{RMSE on time axis results on the Auditory and Visual Oddball EEG-fMRI Test Set.}
    \begin{tabular}{p{2.2cm} p{1.7cm} p{1.7cm} p{1.7cm} p{1.7cm} p{1.7cm}}
        \hline
        Missing & $10\%$ & $25\%$ & $50\%$ & $75\%$ & $90\%$ \\
        \hline
        kNN & $1.21\pm0.86$ & $1.21\pm0.84$ & $1.19\pm0.81$ & $1.11\pm0.71$ & $1.05\pm0.64$ \\
		Barycenter\cite{petitjean2011barycenter} & $1.23\pm0.88$ & $1.25\pm0.87$ & $1.26\pm0.85$ & $1.23\pm0.80$ & $1.22\pm0.77$ \\
		Mean & $1.21\pm0.84$ & $1.17\pm0.80$ & $1.14\pm0.77$ & $\mathbf{1.05}\pm0.66$ & $1.01\pm0.61$ \\
		CE \cite{Pathak_2016_CVPR} & $\mathbf{1.09}\pm0.66$ & $\mathbf{1.08}\pm0.66$ & $\mathbf{1.03}\pm0.62$ & $\mathbf{1.03}\pm0.62$ & $\mathbf{1.00}\pm0.60$ \\
		MICE\cite{white2011multiple} & $1.24\pm0.86$ & $1.27\pm0.88$ & $1.28\pm0.86$ & $3.33\pm3.57$ & $16.97\pm21$ \\
		Dropout & $1.23\pm0.86$ & $1.22\pm0.84$ & $1.21\pm0.82$ & $1.11\pm0.71$ & $1.04\pm0.64$ \\
		Dropout$+D$ & $\mathbf{0.72}\pm0.44$ & $\mathbf{0.81}\pm0.50$ & $\mathbf{0.90}\pm0.57$ & $\mathbf{0.93}\pm0.58$ & $1.06\pm0.65$ \\
		$\Phi$ & $\mathbf{1.23}\pm0.87$ & $\mathbf{1.22}\pm0.84$ & $\mathbf{1.18}\pm0.80$ & $\mathbf{1.06}\pm0.67$ & $\mathbf{1.02}\pm0.61$ \\
		$\Phi+D$ & $\mathbf{0.70}\pm0.43$ & $\mathbf{0.80}\pm0.49$ & $\mathbf{0.87}\pm0.55$ & $\mathbf{0.98}\pm0.58$ & $\mathbf{0.99}\pm0.58$ \\
        \hline
    \end{tabular}
    \label{table:second_rmse_results_time}
\end{table}

\noindent\textbf{Final remarks.}
The rich spatiotemporal nature of neuroimaging modalities such as fMRI, together with their high susceptibility to noise artifacts, create unique difficulties for missing data imputation for both resting-state and stimuli-induced settings \cite{birn2012role}. 
The gathered results indicate that, on one hand, 
no state-of-the-art method for spatial-based imputation stands out, on the other hand, methods able to preserve temporal dependencies, such as DTW-based barycenter, are unable to explore the available data. 
One might argue that if the dataset contains multiple individuals and not a single subject on the same scanner, it might be difficult to fit the correlation matrix, due to different line ups and brain sizes. We recognize it as an obstacle to fit the voxel correlation matrix. To tackle it, we downsample the fMRI spatial resolution by a factor of $6$, going from $29000$ voxels in total to $100$ voxels to represent the whole brain of multiple subjects. The results in Section \ref{section:experiments}, actually show the feasibility of the task and support the claim.

This work shows the clear role of combining time regularization, $D$, with expedite spatial imputation methods to achieve significant improvements on data settings with variable amount and types of missing data. 

We further presented a chained imputation method applied in a neural network setting which achieves state-of-the-art results. 
In fMRI stimuli induced settings, 
the importance of iterative imputation of missing time series in accordance with their pre-computed priority is highlighted by the stable performance of the $\Phi$ layer, even when the missing rate increases.

Missing values imputation, in the perspective of this work, is seen as retrieving information from the no-missing data. Performing imputation all at once is illustrated by the Dropout baseline. $\Phi$ has the advantage of leveraging information from the already imputed information, something Dropout does not do and shows the upside of the proposed approach. In sum, the proposed approach, $\Phi+D$, is competitive and superior to the baselines considered. $\Phi$ shows robustness as the missing rate increases and $D$ is able to remove the spatial prediction noise from $\Phi$ and Dropout, thus being denoted as a denoiser. 
This stable performance is particularly interesting given the heightened differences between resting state and stimuli based fMRI \cite{wehrl2013simultaneous}. 
As future work, we intend to extend this imputation approach to help in modality transfer. The complexity of devising end-to-end approaches for image synthesis from modalities with different spatiotemporal resolutions can be guided under the proposed imputation principles.




\bibliographystyle{unsrtnat}
\bibliography{bibtex}

\newpage
\appendix
\section{Spatial Results}\label{appendix:spatial}

The metrics presented in this Section are the same of Section \ref{subsection:evaluation_metrics}, but instead of being taken along the time axis, they are taken along the spatial/feature axis.

\begin{table}[ht]
\small
    \centering
    \caption{\small MAE on feature axis results on the EEG, fMRI and NODDI Test Set.}
    \begin{tabular}{p{2.2cm} p{1.7cm} p{1.7cm} p{1.7cm} p{1.7cm} p{1.7cm}}
        \hline
        Missing & $10\%$ & $25\%$ & $50\%$ & $75\%$ & $90\%$ \\
        \hline
        kNN & $\mathbf{0.46}\pm1.48$ & $\mathbf{0.44}\pm0.98$ & $\mathbf{0.49}\pm0.85$ & $\mathbf{0.56}\pm0.83$ & $\mathbf{0.62}\pm0.88$ \\
		Barycenter\cite{petitjean2011barycenter} & $0.69\pm2.13$ & $0.73\pm1.51$ & $0.75\pm1.12$ & $0.74\pm0.97$ & $0.75\pm0.92$ \\
		Mean & $0.70\pm2.14$ & $0.64\pm1.29$ & $0.64\pm0.99$ & $0.64\pm0.86$ & $0.64\pm0.82$ \\
		CE \cite{Pathak_2016_CVPR} & $6.59\pm17$ & $2.76\pm4.29$ & $1.48\pm1.45$ & $1.01\pm0.87$ & $0.86\pm0.75$ \\
		MICE\cite{white2011multiple} & $0.59\pm1.78$ & $0.61\pm1.20$ & $0.72\pm1.04$ & $1.39\pm4.63$ & $0.72\pm0.90$ \\
		Dropout & $0.60\pm1.85$ & $0.58\pm1.17$ & $0.60\pm0.88$ & $0.59\pm0.78$ & $0.60\pm0.79$ \\
		Dropout$+D$ & $0.60\pm0.62$ & $0.64\pm0.72$ & $0.65\pm0.89$ & $0.56\pm0.71$ & $0.67\pm0.77$ \\
		$Phi$ & $0.63\pm1.91$ & $0.57\pm1.15$ & $0.59\pm0.87$ & $0.61\pm0.84$ & $0.64\pm0.82$ \\
		$Phi+D$ & $0.58\pm0.60$ & $0.63\pm0.72$ & $0.72\pm0.91$ & $0.59\pm0.81$ & $0.73\pm0.86$ \\
        \hline
    \end{tabular}
    \vspace{-0.5cm}
    \label{table:mae_results}
\end{table}

\begin{table}[ht]
\small
    \centering
    \caption{\small RMSE on feature axis results on the EEG, fMRI and NODDI Test Set.}
    \begin{tabular}{p{2.2cm} p{1.7cm} p{1.7cm} p{1.7cm} p{1.7cm} p{1.7cm}}
        \hline
        Missing & $10\%$ & $25\%$ & $50\%$ & $75\%$ & $90\%$ \\
        \hline
        kNN & $\mathbf{0.82}\pm2.45$ & $\mathbf{0.77}\pm1.49$ & $\mathbf{0.85}\pm1.16$ & $\mathbf{0.95}\pm1.02$ & $\mathbf{1.06}\pm1.03$ \\
		Barycenter\cite{petitjean2011barycenter} & $1.08\pm3.22$ & $1.13\pm2.13$ & $1.12\pm1.43$ & $1.14\pm1.16$ & $1.16\pm1.05$ \\
		Mean & $1.03\pm3.06$ & $0.95\pm1.77$ & $0.99\pm1.28$ & $1.00\pm1.03$ & $1.02\pm0.94$ \\
		CE \cite{Pathak_2016_CVPR} & $3.23\pm7.94$ & $2.04\pm3.01$ & $1.46\pm1.43$ & $1.20\pm0.97$ & $1.11\pm0.82$ \\
		MICE\cite{white2011multiple} & $0.83\pm2.45$ & $0.86\pm1.60$ & $1.05\pm1.33$ & $4.77\pm6.30$ & $1.13\pm1.03$ \\
		Dropout & $0.89\pm2.65$ & $0.85\pm1.60$ & $0.88\pm1.13$ & $0.92\pm0.94$ & $0.97\pm0.91$ \\
		Dropout$+D$ & $0.86\pm0.67$ & $0.96\pm0.79$ & $1.10\pm1.00$ & $0.91\pm0.79$ & $1.02\pm0.85$ \\
		$Phi$ & $0.91\pm2.71$ & $0.84\pm1.56$ & $0.87\pm1.12$ & $0.98\pm1.02$ & $1.02\pm0.94$ \\
		$Phi+D$ & $0.84\pm0.66$ & $0.96\pm0.79$ & $1.16\pm1.01$ & $1.00\pm0.90$ & $1.13\pm0.95$ \\
        \hline
    \end{tabular}
    \vspace{-0.5cm}
    \label{table:rmse_results}
\end{table}

\begin{table}[ht]
\small
    \centering
    \caption{\small MAE on feature axis results on the Auditory and Visual Oddball EEG-fMRI Test Set.}
    \begin{tabular}{p{2.2cm} p{1.7cm} p{1.7cm} p{1.7cm} p{1.7cm} p{1.7cm}}
        \hline
        Missing & $10\%$ & $25\%$ & $50\%$ & $75\%$ & $90\%$ \\
        \hline
        kNN & $0.57\pm1.70$ & $0.60\pm1.13$ & $0.66\pm0.83$ & $0.74\pm0.69$ & $0.86\pm0.65$ \\
		Barycenter\cite{petitjean2011barycenter} & $1.00\pm2.97$ & $1.01\pm1.89$ & $1.01\pm1.24$ & $0.99\pm0.94$ & $0.99\pm0.81$ \\
		Mean & $0.81\pm2.38$ & $0.82\pm1.50$ & $0.81\pm0.98$ & $0.82\pm0.73$ & $0.82\pm0.61$ \\
		CE \cite{Pathak_2016_CVPR} & $7.07\pm19.23$ & $3.02\pm4.59$ & $1.58\pm1.37$ & $1.07\pm0.74$ & $0.92\pm0.60$ \\
		MICE\cite{white2011multiple} & $0.65\pm1.94$ & $0.71\pm1.33$ & $0.78\pm0.99$ & $1.45\pm3.26$ & $3.65\pm16.69$ \\
		Dropout & $0.70\pm2.07$ & $0.69\pm1.29$ & $0.70\pm0.86$ & $0.75\pm0.69$ & $0.81\pm0.62$ \\
		Dropout$+D$ & $\mathbf{0.30}\pm0.44$ & $\mathbf{0.32}\pm0.46$ & $\mathbf{0.39}\pm0.55$ & $\mathbf{0.36}\pm0.52$ & $\mathbf{0.43}\pm0.64$ \\
		$Phi$ & $0.69\pm2.05$ & $0.68\pm1.26$ & $0.71\pm0.87$ & $0.81\pm0.72$ & $0.82\pm0.62$ \\
		$Phi+D$ & $\mathbf{0.28}\pm0.41$ & $\mathbf{0.29}\pm0.40$ & $\mathbf{0.37}\pm0.55$ & $\mathbf{0.43}\pm0.59$ & $\mathbf{0.43}\pm0.64$ \\
        \hline
    \end{tabular}
    \vspace{-0.5cm}
    \label{table:second_mae_results}
\end{table}

\begin{table}[h!]
\small
    \centering
    \caption{\small RMSE on feature axis results on the Auditory and Visual Oddball EEG-fMRI Test Set.}
    \begin{tabular}{p{2.2cm} p{1.7cm} p{1.7cm} p{1.7cm} p{1.7cm} p{1.7cm}}
        \hline
        Missing & $10\%$ & $25\%$ & $50\%$ & $75\%$ & $90\%$ \\
        \hline
        kNN & $0.74\pm2.17$ & $0.76\pm1.39$ & $0.83\pm0.99$ & $0.92\pm0.78$ & $1.04\pm0.70$ \\
		Barycenter\cite{petitjean2011barycenter} & $1.27\pm3.73$ & $1.27\pm2.30$ & $1.25\pm1.46$ & $1.24\pm1.06$ & $1.23\pm0.88$ \\
		Mean & $0.98\pm2.86$ & $0.98\pm1.77$ & $0.98\pm1.13$ & $0.99\pm0.81$ & $0.99\pm0.67$ \\
		CE \cite{Pathak_2016_CVPR} & $2.98\pm6.97$ & $1.91\pm2.52$ & $1.38\pm1.16$ & $1.14\pm0.77$ & $1.05\pm0.64$ \\
		MICE\cite{white2011multiple} & $0.82\pm2.41$ & $0.89\pm1.61$ & $0.99\pm1.18$ & $3.47\pm4.24$ & $17.04\pm22$ \\
		Dropout & $0.89\pm2.61$ & $0.87\pm1.59$ & $0.86\pm1.01$ & $0.92\pm0.77$ & $0.98\pm0.67$ \\
		Dropout$+D$ & $\mathbf{0.54}\pm0.50$ & $\mathbf{0.56}\pm0.51$ & $\mathbf{0.67}\pm0.62$ & $\mathbf{0.63}\pm0.59$ & $\mathbf{0.77}\pm0.72$ \\
		$Phi$ & $0.88\pm2.59$ & $0.85\pm1.54$ & $0.87\pm1.02$ & $0.98\pm0.80$ & $0.99\pm0.67$ \\
		$Phi+D$ & $\mathbf{0.50}\pm0.46$ & $\mathbf{0.50}\pm0.46$ & $\mathbf{0.66}\pm0.62$ & $\mathbf{0.73}\pm0.67$ & $\mathbf{0.77}\pm0.72$ \\
        \hline
    \end{tabular}
    \label{table:second_rmse_results}
\end{table}

By looking at Tables \ref{table:mae_results} and \ref{table:rmse_results}, the extension of a Denoiser, $D$, does not show to be an advantage. Instead, leaving Dropout and $\Phi$, alone have a better performance. CE has a poor performance on the EEG, fMRI and NODDI dataset. kNN shows a good performance spatially, which was expected due to the high neighbour spatial correlation. Although, $D$ does not show superiority in the spatial oriented metrics, it still shows robustness. Further, superiority is clear in Tables \ref{table:second_mae_results} and \ref{table:second_rmse_results} for $D$, we do not go to deep into the reason why this happens, but it may have to do with the different nature of the two datasets (resting state and stimuli based). In sum, $D$ shows the best results in the Auditory and Visual Oddball EEG-fMRI dataset, but $k$NN shows the best results in EEG, fMRI and NODDI dataset (all spatial wise).

\newpage
\section{Random Removal - Results}\label{appendix:random}

\subsection{Random Value Removal}\label{appendix:random_value_removal}

In this Section, we present results by randonly removing values. In contrast, with the results shown in Section \ref{section:discussion}, in which values were removed by region, that is random region removal, here we only remove random values. Spatial location is not considered in this type of removal scheme.

\subsection{Results}\label{appendix:results_random}

\begin{table}[ht]
\small
    \centering
    \caption{\small MAE on spatial axis results on the EEG, fMRI and NODDI Test Set.}
    \begin{tabular}{p{2.2cm} p{1.7cm} p{1.7cm} p{1.7cm} p{1.7cm} p{1.7cm}}
        \hline
        Missing & $10\%$ & $25\%$ & $50\%$ & $75\%$ & $90\%$ \\
        \hline
        kNN & $\mathbf{0.44}\pm1.47$ & $\mathbf{0.47}\pm1.04$ & $\mathbf{0.49}\pm0.84$ & $\mathbf{0.56}\pm0.84$ & $\mathbf{0.67}\pm0.85$ \\
		Barycenter\cite{petitjean2011barycenter} & $0.78\pm2.48$ & $0.77\pm1.57$ & $0.78\pm1.13$ & $0.76\pm0.95$ & $0.77\pm0.92$ \\
		Mean & $0.77\pm2.45$ & $0.74\pm1.52$ & $0.68\pm1.04$ & $0.66\pm0.88$ & $0.64\pm0.81$ \\
		CE\cite{Pathak_2016_CVPR} & $6.99\pm20.02$ & $2.66\pm4.25$ & $1.43\pm1.42$ & $1.03\pm0.85$ & $0.85\pm0.74$ \\
		MICE\cite{white2011multiple} & $0.51\pm1.59$ & $0.53\pm1.06$ & $0.59\pm0.83$ & $0.69\pm0.81$ & $2.67\pm24$ \\
		Dropout & $0.57\pm1.81$ & $0.55\pm1.11$ & $0.54\pm0.77$ & $0.59\pm0.81$ & $0.66\pm0.80$ \\
		Dropout$+D$ & $0.54\pm0.59$ & $0.52\pm0.57$ & $0.61\pm0.69$ & $0.58\pm0.55$ & $0.66\pm0.67$ \\
		$Phi$ & $0.56\pm1.78$ & $0.53\pm1.07$ & $0.52\pm0.80$ & $0.59\pm0.84$ & $0.63\pm0.80$ \\
		$Phi+D$ & $0.54\pm0.60$ & $0.48\pm0.56$ & $0.59\pm0.69$ & $0.56\pm0.61$ & $0.60\pm0.67$ \\
        \hline
    \end{tabular}
    \vspace{-0.5cm}
    \label{table:first_dataset_mae_spatial_results_random}
\end{table}

\begin{table}[ht]
\small
    \centering
    \caption{\small RMSE on spatial axis results on the EEG, fMRI and NODDI Test Set.}
    \begin{tabular}{p{2.2cm} p{1.7cm} p{1.7cm} p{1.7cm} p{1.7cm} p{1.7cm}}
        \hline
        Missing & $10\%$ & $25\%$ & $50\%$ & $75\%$ & $90\%$ \\
        \hline
        kNN & $\mathbf{0.78}\pm2.44$ & $\mathbf{0.81}\pm1.58$ & $\mathbf{0.84}\pm1.14$ & $\mathbf{0.96}\pm1.03$ & $\mathbf{1.06}\pm0.97$ \\
		Barycenter\cite{petitjean2011barycenter} & $1.12\pm3.46$ & $1.12\pm2.14$ & $1.13\pm1.44$ & $1.14\pm1.13$ & $1.17\pm1.04$ \\
		Mean & $1.14\pm3.56$ & $1.11\pm2.12$ & $1.04\pm1.35$ & $1.03\pm1.06$ & $1.01\pm0.92$ \\
		CE\cite{Pathak_2016_CVPR} & $3.25\pm8.40$ & $1.96\pm2.97$ & $1.42\pm1.42$ & $1.19\pm0.94$ & $1.09\pm0.81$ \\
		MICE\cite{white2011multiple} & $0.69\pm2.14$ & $0.75\pm1.42$ & $0.83\pm1.04$ & $0.99\pm0.95$ & $24.91\pm34$ \\
		Dropout & $0.81\pm2.50$ & $0.79\pm1.49$ & $0.77\pm0.98$ & $0.94\pm0.98$ & $1.01\pm0.91$ \\
		Dropout$+D$ & $0.81\pm0.65$ & $0.77\pm0.62$ & $0.92\pm0.76$ & $0.80\pm0.60$ & $0.94\pm0.73$ \\
		$Phi$ & $0.80\pm2.47$ & $0.77\pm1.46$ & $0.80\pm1.04$ & $0.97\pm1.02$ & $1.00\pm0.92$ \\
		$Phi+D$ & $0.81\pm0.66$ & $0.74\pm0.61$ & $0.90\pm0.75$ & $0.83\pm0.67$ & $0.90\pm0.73$ \\
        \hline
    \end{tabular}
    \vspace{-0.5cm}
    \label{table:first_dataset_rmse_spatial_results_random}
\end{table}

\begin{table}[ht]
\small
    \centering
    \caption{\small MAE on time axis results on the EEG, fMRI and NODDI Test Set.}
    \begin{tabular}{p{2.2cm} p{1.7cm} p{1.7cm} p{1.7cm} p{1.7cm} p{1.7cm}}
        \hline
        Missing & $10\%$ & $25\%$ & $50\%$ & $75\%$ & $90\%$ \\
        \hline
        kNN & $0.50\pm0.62$ & $0.52\pm0.65$ & $0.52\pm0.64$ & $0.56\pm0.73$ & $0.66\pm0.80$ \\
		Barycenter\cite{petitjean2011barycenter} & $0.53\pm0.64$ & $0.57\pm0.70$ & $0.63\pm0.70$ & $0.70\pm0.79$ & $0.75\pm0.86$ \\
		Mean & $0.53\pm0.65$ & $0.57\pm0.72$ & $0.59\pm0.70$ & $0.61\pm0.75$ & $0.63\pm0.77$ \\
		CE\cite{Pathak_2016_CVPR} & $0.72\pm0.80$ & $0.68\pm0.74$ & $0.71\pm0.69$ & $0.77\pm0.68$ & $0.77\pm0.70$ \\
		MICE\cite{white2011multiple} & $0.51\pm0.60$ & $0.53\pm0.61$ & $0.57\pm0.61$ & $0.65\pm0.68$ & $2.51\pm24$ \\
		Dropout & $0.51\pm0.60$ & $0.53\pm0.62$ & $0.54\pm0.59$ & $0.58\pm0.70$ & $0.64\pm0.76$ \\
		Dropout$+D$ & $\mathbf{0.51}\pm0.53$ & $\mathbf{0.53}\pm0.56$ & $\mathbf{0.56}\pm0.56$ & $\mathbf{0.59}\pm0.61$ & $\mathbf{0.66}\pm0.76$ \\
		$Phi$ & $0.51\pm0.60$ & $0.52\pm0.62$ & $0.53\pm0.61$ & $0.57\pm0.73$ & $0.62\pm0.76$ \\
		$Phi+D$ & $\mathbf{0.51}\pm0.54$ & $\mathbf{0.51}\pm0.56$ & $\mathbf{0.53}\pm0.56$ & $\mathbf{0.60}\pm0.76$ & $\mathbf{0.64}\pm0.78$ \\
        \hline
    \end{tabular}
    \label{table:first_dataset_mae_time_results_random}
\end{table}

\begin{table}[ht]
\small
    \centering
    \caption{\small RMSE on time axis results on the EEG, fMRI and NODDI Test Set.}
    \begin{tabular}{p{2.2cm} p{1.7cm} p{1.7cm} p{1.7cm} p{1.7cm} p{1.7cm}}
        \hline
        Missing & $10\%$ & $25\%$ & $50\%$ & $75\%$ & $90\%$ \\
        \hline
        kNN & $0.79\pm0.68$ & $0.83\pm0.73$ & $0.83\pm0.71$ & $0.92\pm0.81$ & $1.04\pm0.88$ \\
		Barycenter\cite{petitjean2011barycenter} & $0.83\pm0.70$ & $0.90\pm0.77$ & $0.94\pm0.76$ & $1.06\pm0.86$ & $1.14\pm0.94$ \\
		Mean & $0.84\pm0.72$ & $0.92\pm0.80$ & $0.92\pm0.77$ & $0.97\pm0.83$ & $0.99\pm0.85$ \\
		CE\cite{Pathak_2016_CVPR} & $1.08\pm0.87$ & $1.00\pm0.80$ & $0.99\pm0.74$ & $1.03\pm0.73$ & $1.04\pm0.75$ \\
		MICE\cite{white2011multiple} & $0.79\pm0.66$ & $0.81\pm0.67$ & $0.84\pm0.67$ & $0.95\pm0.74$ & $24.79\pm33$ \\
		Dropout & $0.79\pm0.66$ & $0.81\pm0.68$ & $0.80\pm0.65$ & $0.91\pm0.77$ & $0.99\pm0.84$ \\
		Dropout$+D$ & $\mathbf{0.74}\pm0.58$ & $\mathbf{0.76}\pm0.61$ & $\mathbf{0.79}\pm0.60$ & $\mathbf{0.85}\pm0.67$ & $\mathbf{1.01}\pm0.83$ \\
		$Phi$ & $0.79\pm0.67$ & $0.81\pm0.68$ & $0.81\pm0.67$ & $0.92\pm0.81$ & $0.98\pm0.85$ \\
		$Phi+D$ & $\mathbf{0.75}\pm0.59$ & $\mathbf{0.76}\pm0.61$ & $\mathbf{0.77}\pm0.61$ & $\mathbf{0.98}\pm0.85$ & $\mathbf{1.00}\pm0.86$ \\
        \hline
    \end{tabular}
    \vspace{-0.5cm}
    \label{table:first_dataset_rmse_time_results_random}
\end{table}

\begin{table}[ht]
\small
    \centering
    \caption{\small MAE on spatial axis results on the Auditory and Visual Oddball EEG-fMRI Test Set.}
    \begin{tabular}{p{2.2cm} p{1.7cm} p{1.7cm} p{1.7cm} p{1.7cm} p{1.7cm}}
        \hline
        Missing & $10\%$ & $25\%$ & $50\%$ & $75\%$ & $90\%$ \\
        \hline
        kNN & $0.59\pm1.83$ & $0.62\pm1.18$ & $0.66\pm0.83$ & $0.75\pm0.70$ & $0.87\pm0.67$ \\
		Barycenter\cite{petitjean2011barycenter} & $1.07\pm3.30$ & $1.04\pm1.95$ & $1.01\pm1.26$ & $1.00\pm0.94$ & $1.00\pm0.81$ \\
		Mean & $0.82\pm2.53$ & $0.83\pm1.54$ & $0.81\pm0.99$ & $0.82\pm0.73$ & $0.82\pm0.63$ \\
		CE\cite{Pathak_2016_CVPR} & $0.66\pm2.05$ & $0.70\pm1.35$ & $0.73\pm0.94$ & $0.80\pm0.77$ & $1.62\pm5.24$ \\
		MICE\cite{white2011multiple} & $0.74\pm2.29$ & $0.71\pm1.34$ & $0.69\pm0.87$ & $0.77\pm0.71$ & $0.81\pm0.63$ \\
		Dropout & $\mathbf{0.30}\pm0.45$ & $\mathbf{0.33}\pm0.47$ & $\mathbf{0.31}\pm0.46$ & $\mathbf{0.39}\pm0.55$ & $\mathbf{0.43}\pm0.62$ \\
		Dropout$+D$ & $0.69\pm2.15$ & $0.67\pm1.28$ & $0.68\pm0.86$ & $0.80\pm0.72$ & $0.83\pm0.63$ \\
		$Phi$ & $\mathbf{0.29}\pm0.45$ & $\mathbf{0.33}\pm0.46$ & $\mathbf{0.31}\pm0.45$ & $\mathbf{0.43}\pm0.59$ & $\mathbf{0.43}\pm0.60$ \\
		$Phi+D$ & $0.58\pm0.60$ & $0.63\pm0.72$ & $0.72\pm0.91$ & $0.59\pm0.81$ & $0.73\pm0.86$ \\
        \hline
    \end{tabular}
    \vspace{-0.5cm}
    \label{table:second_dataset_mae_spatial_results_random}
\end{table}

\begin{table}[ht]
\small
    \centering
    \caption{\small RMSE on spatial axis results on the Auditory and Visual Oddball EEG-fMRI Test Set.}
    \begin{tabular}{p{2.2cm} p{1.7cm} p{1.7cm} p{1.7cm} p{1.7cm} p{1.7cm}}
        \hline
        Missing & $10\%$ & $25\%$ & $50\%$ & $75\%$ & $90\%$ \\
        \hline
        kNN & $0.76\pm2.34$ & $0.79\pm1.46$ & $0.83\pm1.00$ & $0.93\pm0.80$ & $1.06\pm0.73$ \\
		Barycenter\cite{petitjean2011barycenter} & $1.32\pm4.03$ & $1.28\pm2.35$ & $1.26\pm1.48$ & $1.24\pm1.05$ & $1.24\pm0.88$ \\
		Mean & $1.00\pm3.05$ & $1.00\pm1.83$ & $0.99\pm1.15$ & $0.99\pm0.82$ & $1.00\pm0.68$ \\
		CE\cite{Pathak_2016_CVPR} & $0.86\pm2.64$ & $0.91\pm1.69$ & $0.94\pm1.13$ & $1.01\pm0.87$ & $5.45\pm6.72$ \\
		MICE\cite{white2011multiple} & $0.92\pm2.83$ & $0.89\pm1.65$ & $0.87\pm1.03$ & $0.95\pm0.80$ & $0.99\pm0.69$ \\
		Dropout & $\mathbf{0.54}\pm0.52$ & $\mathbf{0.58}\pm0.53$ & $\mathbf{0.55}\pm0.52$ & $\mathbf{0.68}\pm0.62$ & $\mathbf{0.76}\pm0.70$ \\
		Dropout$+D$ & $0.87\pm2.66$ & $0.85\pm1.57$ & $0.86\pm1.01$ & $0.98\pm0.81$ & $1.00\pm0.68$ \\
		$Phi$ & $\mathbf{0.53}\pm0.51$ & $\mathbf{0.57}\pm0.52$ & $\mathbf{0.55}\pm0.51$ & $\mathbf{0.73}\pm0.66$ & $\mathbf{0.73}\pm0.67$ \\
		$Phi+D$ & $0.58\pm0.60$ & $0.63\pm0.72$ & $0.72\pm0.91$ & $0.59\pm0.81$ & $0.73\pm0.86$ \\
        \hline
    \end{tabular}
    \vspace{-0.5cm}
    \label{table:second_dataset_rmse_spatial_results_random}
\end{table}

\begin{table}[ht]
\small
    \centering
    \caption{\small MAE on time axis results on the Auditory and Visual Oddball EEG-fMRI Test Set.}
    \begin{tabular}{p{2.2cm} p{1.7cm} p{1.7cm} p{1.7cm} p{1.7cm} p{1.7cm}}
        \hline
        Missing & $10\%$ & $25\%$ & $50\%$ & $75\%$ & $90\%$ \\
        \hline
        kNN & $0.91\pm0.82$ & $0.94\pm0.82$ & $0.92\pm0.78$ & $0.89\pm0.70$ & $0.87\pm0.64$ \\
		Barycenter\cite{petitjean2011barycenter} & $0.92\pm0.83$ & $0.96\pm0.83$ & $0.96\pm0.79$ & $0.98\pm0.77$ & $0.98\pm0.74$ \\
		Mean & $0.90\pm0.81$ & $0.92\pm0.80$ & $0.89\pm0.75$ & $0.85\pm0.66$ & $0.84\pm0.60$ \\
		CE\cite{Pathak_2016_CVPR} & $0.93\pm0.84$ & $0.97\pm0.85$ & $0.98\pm0.83$ & $0.91\pm0.73$ & $1.57\pm5.12$ \\
		MICE\cite{white2011multiple} & $0.92\pm0.83$ & $0.95\pm0.82$ & $0.94\pm0.79$ & $0.88\pm0.69$ & $0.85\pm0.62$ \\
		Dropout & $0.67\pm0.52$ & $0.66\pm0.53$ & $0.69\pm0.53$ & $0.72\pm0.56$ & $0.78\pm0.56$ \\
		Dropout$+D$ & $0.92\pm0.83$ & $0.95\pm0.82$ & $0.93\pm0.78$ & $0.86\pm0.67$ & $0.84\pm0.61$ \\
		$Phi$ & $0.67\pm0.51$ & $0.66\pm0.52$ & $0.67\pm0.52$ & $0.83\pm0.57$ & $0.82\pm0.56$ \\
		$Phi+D$ & $\mathbf{0.58}\pm0.60$ & $\mathbf{0.63}\pm0.72$ & $\mathbf{0.72}\pm0.91$ & $\mathbf{0.59}\pm0.81$ & $\mathbf{0.73}\pm0.86$ \\
        \hline
    \end{tabular}
    \vspace{-0.5cm}
    \label{table:second_dataset_mae_time_results_random}
\end{table}

\begin{table}[ht]
\small
    \centering
    \caption{\small RMSE on time axis results on the Auditory and Visual Oddball EEG-fMRI Test Set.}
    \begin{tabular}{p{2.2cm} p{1.7cm} p{1.7cm} p{1.7cm} p{1.7cm} p{1.7cm}}
        \hline
        Missing & $10\%$ & $25\%$ & $50\%$ & $75\%$ & $90\%$ \\
        \hline
        kNN & $1.23\pm0.88$ & $1.25\pm0.88$ & $1.20\pm0.83$ & $1.14\pm0.74$ & $1.08\pm0.67$ \\
		Barycenter\cite{petitjean2011barycenter} & $1.23\pm0.89$ & $1.27\pm0.89$ & $1.25\pm0.84$ & $1.24\pm0.81$ & $1.23\pm0.78$ \\
		Mean & $1.22\pm0.87$ & $1.22\pm0.85$ & $1.16\pm0.79$ & $1.08\pm0.70$ & $1.03\pm0.63$ \\
		CE\cite{Pathak_2016_CVPR} & $1.25\pm0.90$ & $1.29\pm0.91$ & $1.29\pm0.89$ & $1.17\pm0.78$ & $5.36\pm6.37$ \\
		MICE\cite{white2011multiple} & $1.23\pm0.89$ & $1.26\pm0.88$ & $1.23\pm0.84$ & $1.12\pm0.73$ & $1.05\pm0.65$ \\
		Dropout & $0.85\pm0.56$ & $0.85\pm0.56$ & $0.87\pm0.56$ & $0.91\pm0.59$ & $0.96\pm0.59$ \\
		Dropout$+D$ & $1.24\pm0.89$ & $1.26\pm0.88$ & $1.21\pm0.83$ & $1.09\pm0.71$ & $1.04\pm0.64$ \\
		$Phi$ & $0.84\pm0.54$ & $0.84\pm0.55$ & $0.85\pm0.55$ & $1.00\pm0.59$ & $1.00\pm0.59$ \\
		$Phi+D$ & $\mathbf{0.58}\pm0.60$ & $\mathbf{0.63}\pm0.72$ & $\mathbf{0.72}\pm0.91$ & $\mathbf{0.59}\pm0.81$ & $\mathbf{0.73}\pm0.86$ \\
        \hline
    \end{tabular}
    \label{table:second_dataset_rmse_time_results_random}
\end{table}

Tables \ref{table:first_dataset_mae_spatial_results_random}, \ref{table:first_dataset_rmse_spatial_results_random}, \ref{table:first_dataset_mae_time_results_random} and \ref{table:first_dataset_rmse_time_results_random} show the results for the EEG, fMRI and NODDI dataset. $k$NN shows superiority in the spatial metrics, while the addition of a Denoiser, $D$, does not have a big significance in this dataset.

Tables \ref{table:second_dataset_mae_spatial_results_random}, \ref{table:second_dataset_rmse_spatial_results_random}. \ref{table:second_dataset_mae_time_results_random} and \ref{table:second_dataset_rmse_time_results_random} show the results for the Auditory and Visual Oddball EEG-fMRI datset. $\Phi+D$ shows the best results in the time wise metrics with a significatn superiority shown in Tables \ref{table:second_dataset_mae_time_results_random} and \ref{table:second_dataset_rmse_time_results_random}. On the other hand, $\Phi$ has the best results in the spatial/feature wise results in Tables \ref{table:second_dataset_mae_spatial_results_random} and \ref{table:second_dataset_rmse_spatial_results_random}, which in Appendix \ref{appendix:spatial} was $k$NN the best model. This is due to the different types of missing values settings, here removal is random, but in Appendix \ref{appendix:spatial} removal was made by region, bringing an advantage to neighbour methods, such as $k$NN.

\end{document}